\documentclass[12pt]{iopart}

\pdfoutput=1 

\usepackage{graphicx}
\usepackage{color}
\usepackage{soul}

\begin{document}
\title{The scaling of the minimum sum of edge lengths in uniformly random trees}

\author{Juan Luis Esteban$^1$, Ramon Ferrer-i-Cancho$^2$ and Carlos G\'omez-Rodr\'iguez$^3$}
\address{$^1$ Logic and Programming, LOGPROG Research Group \\
Departament de Ci\`encies de la Computaci\'o, \\
Universitat Polit\`ecnica de Catalunya, \\
Campus Nord, Edifici Omega, Jordi Girona Salgado 1-3. \\
08034 Barcelona, Catalonia (Spain)}
\address{$^2$ Complexity \& Quantitative Linguistics Lab, LARCA Research Group \\
Departament de Ci\`encies de la Computaci\'o, \\
Universitat Polit\`ecnica de Catalunya, \\
Campus Nord, Edifici Omega, Jordi Girona Salgado 1-3. \\
08034 Barcelona, Catalonia (Spain)}
\address{$^3$ LyS Research Group \\ 
Departamento de Computaci\'on, \\ 
Facultade de Inform\'atica, \\ 
Universidade da Coru\~na, \\
Campus de A Coru\~na, 15071 A Coru\~na, Spain
}
\ead{\{esteban,rferrericancho\}@cs.upc.edu, cgomezr@udc.es}
\begin{abstract}
The minimum linear arrangement problem on a network consists of finding the minimum sum of edge lengths that can be achieved when the vertices
are arranged linearly. Although there are algorithms to solve this problem on trees in polynomial time, they have remained theoretical and have not been implemented in practical contexts to our knowledge. Here we use one of those algorithms to investigate the growth of this sum as a function of the size of the tree in uniformly random trees. We show that this sum is bounded above by its value in a star tree. We also show that the mean edge length grows logarithmically in optimal linear arrangements, in stark contrast to the linear growth that is expected on optimal arrangements of star trees or on random linear arrangements.
\end{abstract}




\maketitle

\section{Introduction}

By the end of the last century, research on graphs was revolutionized by a series of discoveries on the statistical properties of many real networks \cite{Watts1998,Redner1998,Barabasi1999}: 
\begin{itemize}
\item
Degree distributions exhibit heavy tails, in stark contrast to the binomial distribution of Erd\H{o}s-R\'enyi graphs \cite{Redner1998,Barabasi1999}.
\item
Their cliquishness, i.e. the probability that the first neighbours of a node are connected, is high 
while in the corresponding Erd\H{o}s-R\'enyi graph this probability is low because it coincides with the network density of links \cite{Watts1998}. 
\item
The so-called small-world phenomenon, i.e. the average geodesic distance between vertices (the average minimum vertex-vertex distance) that is denoted by $\delta$, is low compared to $n$, the number of vertices of the network \cite{Watts1998}. This phenomenon is also shared with Erd\H{o}s-R\'enyi graphs (provided that their density of links is large enough).  
In these graphs, one has \cite{Watts1998}
\begin{equation}
\delta \approx \frac{\log n}{\log \left< k \right>},
\label{small_world_equation}
\end{equation}
where $\left< k \right>$ is the mean degree of vertices ($\left< k \right> \gg \log n$ is needed by Eq. \ref{small_world_equation}). 
A much slower scaling of $\delta$ with respect to $n$ is found in networks with power-law degree distributions \cite{Cohen2002a}. 
\end{itemize}

These seminal works spurred an industry of both theoretical and empirical research (e.g., \cite{Barrat2008a,Newman2010a,Costa2007a} and references therein). One avenue has been the investigation of the networks or ensembles of networks that result from imposing certain constraints over the exponentially huge space of possible networks \cite{Molloy1995a, Roberts2013a}. A fundamental contribution has come from approaches that extend ideas and concepts from statistical mechanics and information theory to complex network ensembles \cite{Roberts2013a,Anand2010a}. A precursor of this approach is the configuration model, which focuses on an ensemble of networks that have the same degree sequence \cite{Molloy1995a}. 
Further examples are research shedding light on the prevalence of disassortative mixing in real networks \cite{Johnson2010a}
or analyses that unveil the higher level of order of network ensembles with power-law degree distributions with respect to networks with homogeneous degree distributions \cite{Bianconi2008a}.

In the investigations reviewed above, the topology of the network is free {\em a priori}. Another possibility is to fix the network topology and impose further constraints on it. This takes us to another research avenue that started before the complex networks revolution: applications of statistical mechanics to solve combinatorial optimization problems \cite{Lai1987a,dAuriac1997a}. Vertex coloring, perhaps one of the most popular of these combinatorial problems \cite{Lai1987a}, consists of assigning numbers from $1$ to $m$ to vertices (every number representing a different color) so that $m$ is minimized under the constraint that no two connected vertices are assigned the same number. A perhaps less popular example is the minimum cut linear arrangement problem (also known as mincut or cutwidth problem) \cite{Diaz2002}, which has also been investigated with statistical mechanics tools \cite{dAuriac1997a}.
The target of this article is another NP-hard optimization problem, a sister of the minimum cut linear arrangement problem, namely the minimum linear arrangement (m.l.a.) problem \cite{Diaz2002}:
 the problem of assigning distinctive integers from $1$ to $n$ to each vertex so as to minimize $D$, defined as the sum of the absolute differences between numbers at both ends of every edge. 
A more detailed definition of the m.l.a. problem will be presented next to introduce notation, indicate further connections with statistical physics and present some specific motivations of our work. 

Suppose that the vertices of a network are sorted in a sequence and that the length of an edge is defined as the distance between 
the vertices involved. The m.l.a. problem consists of finding the minimum sum of edge lengths over all possible orderings of vertices \cite{Diaz2002}.
More formally, suppose that the network has $n$ vertices and that $\pi(v)$ is the position of vertex $v$ in an ordering of the vertices ($1 \leq \pi(v) \leq n$). $\pi$ is a one-to-one mapping between vertices and integers between $1$ and $n$. The sum of edge lengths can be defined as a sum over all edges as
\begin{equation}
D = \sum_{u\sim v} |\pi(u) - \pi(v)|,
\end{equation}  
where $u \sim v$ indicates an edge between vertices $u$ and $v$ and $|\pi(u) - \pi(v)|$ is the length of $u\sim v$.
Solving the m.l.a. problem for a given network consists of finding $D_{min}$, the minimum value of $D$ among all the possible $\pi$ that define a linear ordering of the vertices. The $\pi$'s where $D = D_{min}$ define minimum linear arrangements. 
Although the solution of the m.l.a. problem is an NP-hard optimization problem in general, polynomial time algorithms for undirected trees are available \cite{Goldberg1976a,Shiloach1979,Chung1984}.

Here we will investigate the scaling of $D_{min}$ as $n$ increases over the ensemble of uniformly random trees, where the m.l.a. problem is computationally tractable. As a tree has $n - 1$ edges, $D_{min}/(n-1)$ is the mean length of edges in a minimum linear arrangement. Here we will show that $D_{min}/(n-1)$ grows logarithmically with the size of the random tree, a feature reminiscent of Eq. \ref{small_world_equation} for unrestricted networks. The m.l.a. problem can be seen as a particular case of an arrangement of a tree in an $m$-dimensional lattice, where $m=1$ in the customary formulation of the problem. In this regard, our research is related to studies on spanning trees in $m$-dimensional lattices \cite{Manna1992a,Barthelemy2006a}. While in our case the tree structure is fixed and the goal is to find an optimal ordering of the vertices, the tree structure is variable in those studies. As a problem of constraints on the ensemble of possible permutations of a sequence (defined by the vertices of a tree), the m.l.a. is connected with research on the distribution of the distance between elements in a sequence, with edge length being a particular case \cite{Zornig1984a,Ferrer2004b}. If no constraint is imposed, the probability that an edge has a certain length decays linearly with the distance between the vertices \cite{Zornig1984a,Ferrer2004b}. When $D$ is constrained (not necessarily reaching $D_{min}$), an exponential-like distribution is obtained \cite{Ferrer2004b}. Interestingly, an exponential decay of probability is found in real syntactic dependency trees \cite{Ferrer2004b}.
     
The motivation of our work is three-fold. 

First, we aim to expand a large body of research on the scaling of tree properties as $n$ increases, e.g., \cite{Cayley1889a,Otter1948a,Moon1970a,Noy1998a,Ferrer2014f}. A popular example is the growth of $t(n)$, the number of different trees of $n$ vertices, which is \cite{Cayley1889a} 
\begin{equation}
t(n) = n^{n-2}
\end{equation}
for labelled trees. Concerning unlabelled trees, the calculation of $t(n)$ is a harder problem but it is known that \cite{Otter1948a}
\begin{equation}
t(n) \sim c_1 \alpha c_2^n n^{-5/2}
\end{equation}
as $n \rightarrow \infty$ with $c_1$ and $c_2$ being two constants. 

Another example of scaling law is the expectation of $V[k]$, the degree variance of a tree, in uniformly random trees of a given number of vertices, which obeys \cite{Moon1970a,Noy1998a}
\begin{equation}
\left< V[k]\right> =  \left(1- \frac{1}{n}\right)\left(1-\frac{2}{n}\right).
\end{equation}
Hereafter we use $\left< ... \right>$ to refer to expectations over the ensemble of uniformly random labelled trees with a certain number of vertices $n$. In this article, we will contribute with an investigation of the relationship between $\left<D_{min}\right>$ and $n$.     

Second, $D_{min}$ is a baseline for research on the scaling of $D$ in syntactic dependency trees \cite{Ferrer2004b,Ferrer2013c,Park2009a,Futrell2015a},
and thus the scaling of $D_{min}$ in uniformly random trees could also be a reference or baseline for future research on the scaling of $D_{min}$ in syntactic dependency trees.

Third, algorithms for solving the m.l.a. problem on trees \cite{Shiloach1979,Chung1984} have remained theoretical. As far as we know, they have never been implemented for practical reasons. Implementing them in a less theoretical setup gives us a chance to verify their correctness.  

The remainder of the article is organized as follows. Section \ref{background_section} presents some technical background and definitions that are necessary for other sections.
Section \ref{upper_bound_section} derives an upper bound of $D_{min}$. 
Section \ref{scaling_section} presents the logarithmic growth of $\left<D_{min}\right>/(n-1)$ as a function of $n$ and related results of computer simulations over the ensemble of uniformly random trees. Finally, Section \ref{discussion_section} discusses all the results obtained.

\section{Background}

\label{background_section}

\begin{figure}
\begin{indented}
\item[]
\includegraphics{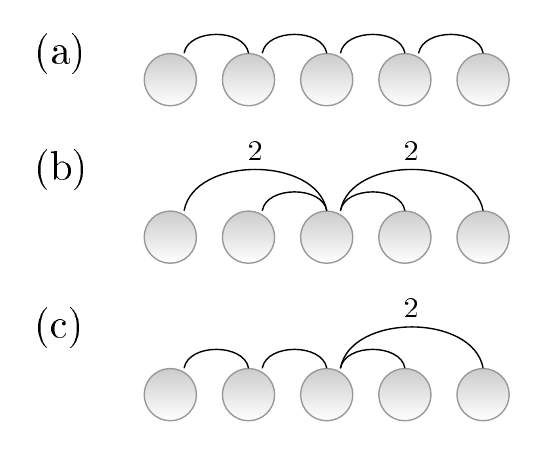}
\caption{\label{trees_with_5_nodes} All (unlabelled) undirected trees with 5 nodes, (a) linear tree, (b) star tree, (c) quasi-star tree \cite{Ferrer2014f}. The lengths of edges that are greater than 1 are indicated.}
\end{indented}
\end{figure}

Analytical solutions for $D_{min}$ are available for certain kinds of trees:
\begin{itemize}
\item
A linear tree (e.g., figure \ref{trees_with_5_nodes} (a)), a tree whose maximum vertex degree is 2 \cite{Ferrer2013d}.   
In a linear tree \cite{Ferrer2013b}, 
   \begin{equation}
   D_{min} = n - 1. 
   \label{mla_linear_tree_equation}
   \end{equation}
\item
A star tree (e.g., figure \ref{trees_with_5_nodes} (b)), a tree with one vertex with maximum degree, the rest with degree 1 \cite{Ferrer2013d}.   
In a star tree \cite{Ferrer2013e},
   \begin{equation} 
   D_{min} = \frac{n^2 - x}{4},
   \label{mla_star_tree_equation}
   \end{equation}
   where $x$ indicates if $n$ is odd ($x = 1$ if $n$ is odd and $x = 0$ otherwise).
\item
In a complete binary tree,  
   \begin{equation}
   D_{min} = 2^k \left(\frac{k}{3} + \frac{5}{18}\right) + (-1)^k \frac{2}{9} - 2,
   \label{mla_complete_binary_tree_equation}
   \end{equation}
where $k = \log_2(n + 1)$ is the number of levels \cite{Chung1978a}.  
\item
In complete trees of $k$ levels where the root is attached to a couple of complete ternary trees of $k-1$ levels,
   \begin{equation}
   D_{min} = 2(k - 1)3^{k - 2} 
   \label{mla_kind_of_ternary_tree_equation} 
   \end{equation}
for $k \geq 2$ \cite{Chung1978a}.
\end{itemize}

It is easy to see that $D_{min} \geq D_{min}^{linear}$, where $D_{min}^{linear}$ is the value of $D_{min}$ of a linear tree with the same number of vertices, defined in equation (\ref{mla_linear_tree_equation}) \cite{Ferrer2013b}.  
Here will show that $D_{min} \leq D_{min}^{star}$, where $D_{min}^{star}$ is the value of $D_{min}$ of a star tree with the same number of vertices, defined in equation (\ref{mla_star_tree_equation}).

$\left< D_{min} \right>$, the expectation of $D_{min}$ in the ensemble of uniformly labelled trees with a certain number of vertices, can also be seen as the average value of $D_{min}$ in all possible labelled trees of the same size. 
Suppose that $\left< D \right>$ is the average value of $D$ in uniformly random trees where vertex labels are taken as vertex positions. Then the growth of $\left<D \right>$ as a function of $n$ should be close to $D_{random}$, the expectation of $D$ in uniformly random linear arrangements of the $n$ vertices of an arbitrary tree, which is \cite{Zornig1984a,Ferrer2004b}
\begin{equation}
D_{random} = \frac{(n-1)(n+1)}{3}.
\label{rla_scaling_of_lengths_equation}
\end{equation}  
Since trees have $n - 1$ edges, their mean edge length is $D/(n-1)$ \cite{Ferrer2004b}.
We will show that $\left< D_{min} \right>/(n-1)$ grows logarithmically for $n \geq 3$, i.e. 
\begin{equation}
\left<D_{min}\right>/(n-1) \approx a \log n + b,
\label{scaling_of_mean_mla_equation}
\end{equation}
where $a$ and $b$ are two constants and then
\begin{equation}
\left<D_{min}\right> \approx a (n-1) \log n + (n - 1) b.
\end{equation}
Note that (\ref{scaling_of_mean_mla_equation}) is in stark contrast to the linear growth of $D/(n-1)$ in uniformly random linear arrangements -- see (\ref{rla_scaling_of_lengths_equation}) -- or the upper bound provided by optimal linear arrangements of star trees -- see (\ref{mla_star_tree_equation}).

The next section presents a derivation of an upper bound of $D_{min}$ that is in turn bounded above by $D_{random}$. 

\section{Upper bound for $D_{min}$}

\label{upper_bound_section}

Suppose an algorithm A to obtain a linear arrangement $\pi$ for a tree $T$ with $n \geq 1$:
\begin{itemize}
\item
If $n = 1$ then $\pi(1) = 1$ and finish the algorithm.
\item
Select a leaf $u$ (every tree where $n > 1$ has at least two leaves \cite[p. 11]{Bollobas1998a}).
\item
Let $T'$ be the result of removing $u$ from $T$.
\item
Obtain a linear arrangement $\pi'$ for $T'$ recursively with this algorithm.
\item
Let us use the subindex $f$ to refer to first and the subindex $l$ to refer to last. Accordingly, 
let $\pi_f$ be the linear arrangement consisting of placing $u$ first ($f$) followed by the remainder of the vertices according to $\pi'$. Similarly, let $\pi_l$ be the linear arrangement consisting of placing $u$ last ($l$) preceded by the remainder of the vertices according to $\pi'$.
Let $v$ be the node to which $u$ is attached in $T$. 
Let $d_f$ be the length of the edge $u \sim v$ in $\pi_f$ and $d_l$ be the length of that edge in $\pi_l$.
\item
Find a linear arrangement for $T$ given $\pi'$: if $d_f < d_l$ then $\pi = \pi_f$; $\pi = \pi_l$ otherwise.
\end{itemize}

\begin{figure}
\begin{indented}
\item[]
\includegraphics{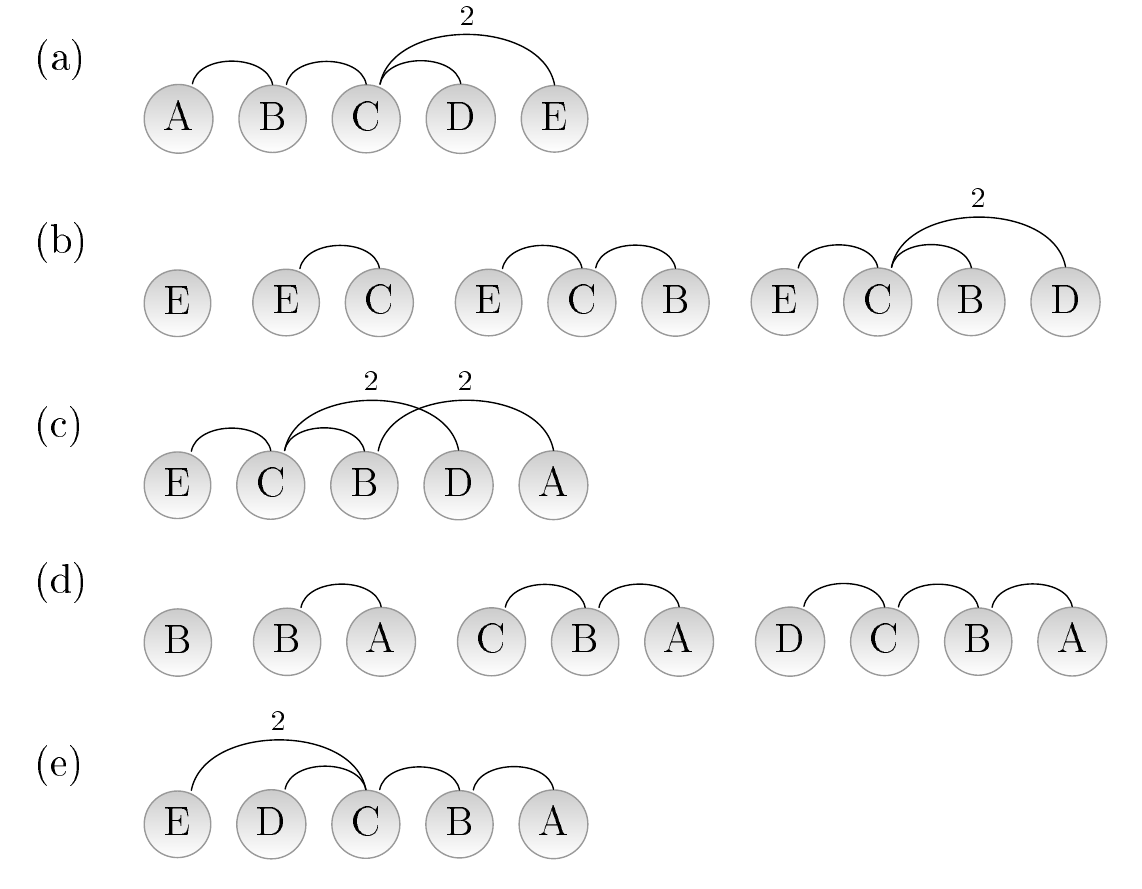} 
\caption{\label{example_algorithm} (a) Minimum linear arrangement of a tree of 5 vertices. (b-c) A sequence of linear arrangements produced by algorithm A for the tree in (a) that ends with a suboptimal arrangement in (c) ($D = 6$). (d-e) Another sequence ending with an optimal arrangement in (e) ($D = 5$). }
\end{indented}
\end{figure}

As an illustration of this algorithm, let us consider figure~\ref{example_algorithm} (a) where there is an optimal linear arrangement of a quasi-star tree with 5 nodes. Figures~\ref{example_algorithm} (b-c) show the series of linear arrangements that the algorithm produces when the leaves that it chooses follow the order A, D, B, C. Notice that the final linear arrangement (figure ~\ref{example_algorithm} (c)) is not optimal (recall figure~\ref{example_algorithm} (a)). Figure~\ref{example_algorithm} (b) shows that the first arrangement has only node E. For the second arrangement the algorithm places node C after the arrangement for node E. For the third arrangement the algorithm places node B after the arrangement for nodes E and C, etc.
In contrast, when the order is E, D, C, A, the algorithm produces a series of linear arrangements (figure ~\ref{example_algorithm} (d-e)) where the final linear arrangement is optimal (figure ~\ref{example_algorithm} (e)).

Let $D_A$ be the sum of dependency lengths of a linear arrangement produced with the linearization algorithm above. By definition, $D_{min} \leq D_A$. 
It is easy to see that $D_{min}=D_A$ for linear trees and star trees of any number of nodes. The question is whether $D_{min} = D_{A}$ in general. As trees for $1 \leq n \leq 4$ are only star trees or linear trees, examples where $D_{min} < D_A$ require $n \geq 5$. Figure~\ref{trees_with_5_nodes} shows all the trees with 5 nodes: a linear tree, a star tree and a quasi-star tree. Since Algorithm A satisfies $D_{min} = D_{A}$ for linear and star trees, we just need to check whether $D_{min} = D_{A}$ or not for that quasi-star tree. It turns out that the algorithm can produce linear arrangements that are not optimal by an unlucky choice of the order of leaves as we have shown in figures~\ref{example_algorithm} (b-c). Similar counterexamples can be built for larger trees.
Therefore, we conclude that $D_{min} \leq D_{A}$ for $n > 4$ and $D_{min} = D_{A}$ for $n \leq 4$. Those familiar with the complexity of efficient algorithms for solving the minimum linear arrangement problem \cite{Shiloach1979,Chung1984} should not find surprising that $D_{min} = D_{A}$ does not hold in general.

Now we will derive an upper bound for $D_A$. If $T$ has one vertex then $D_A = 0$. If $T$ has at least two vertices then 
\begin{equation}
D_A = \min(d_f, d_l) + D_A',
\label{prerecursive_equation}
\end{equation}
where $D_A'$ is the sum of dependency lengths of $T'$. $\min(d_f, d_l)$ can be calculated easily with the help of $\pi'$. Since $d_f = \pi'(v)$ while $d_l = n - \pi'(v)$, one has that
\begin{equation}
\min(d_f, d_l) = \min(\pi'(v), n - \pi'(v)) 
\end{equation}
As $1 \leq \pi'(v) \leq n - 1$,
\begin{equation}
\min(d_f, d_l) \leq \max_{1 \leq \pi'(v) \leq n - 1} \min(\pi'(v), n - \pi'(v)), 
\end{equation}
we get
\begin{equation}
\min(d_f, d_l) \leq 
   \left\{
      \begin{array}{lr}  
      \frac{n}{2} \mbox{~if $n$ is even} \\
      \frac{n - 1}{2} \mbox{~if $n$ is odd}.
      \end{array}
   \right.
   \label{edge_length_upper_bound_equation}
\end{equation}
Intuitively, (\ref{edge_length_upper_bound_equation}) means that the worst case for the minimal length of the edge $u\sim v$ is when $v$ is in the middle of the linear arrangement, so both $d_f$ and $d_l$ are large. If $v$ is, say, near the beginning, $d_l$ will be large, but $d_f$ will be small and therefore the minimum of both will be small.

Knowing this, $D_{A,max}$, an upper bound for $D_A$, is easy to derive assuming $n \geq 1$. 
Suppose that $D_A(n)$ is the sum of dependency lengths produced by algorithm A for a tree of $n$ vertices and $D_{A,max}(n)$ is an upper bound of it. Then equation (\ref{prerecursive_equation}) gives  
\begin{eqnarray}
D_A(n) & =    & \min(d_f, d_l) + D_A(n-1) \\
       & \leq & \min(d_f, d_l) + D_{A,max}(n-1) = D_{A,max}(n) \label{recursive_equation}
\end{eqnarray}
with $D_A(1) = D_{A,max}(1) = 0$.  

If a tree has $n$ vertices it has $n-1$ edges and then Algorithm $A$ produces the length of $n-1$ edges. If $n$ is odd, the recursive application of equation (\ref{recursive_equation}) and the definition of $\min(d_f, d_l)$ in (\ref{edge_length_upper_bound_equation}) give 
\begin{equation}
\fl D_{A,max}(n) = \frac{n - 1}{2} + \frac{n-1}{2}+ \frac{n - 3}{2} +  \frac{n - 3}{2} + \frac{n - 5}{2} + \frac{n - 5}{2} + ... + 2 + 2 + 1 + 1.  
\end{equation} 
Thus, one has
\begin{eqnarray}
D_{A,max} & = & 2 \sum_{i = 1}^{\frac{n - 1}{2}} i \\
          & = & \frac{(n - 1)(n + 1)}{4}. 
\end{eqnarray}
In $n$ is even, the recursive application of equation (\ref{recursive_equation}) and the definition of $\min(d_f, d_l)$ in (\ref{edge_length_upper_bound_equation}) give 
\begin{equation}
\fl D_{A,max}(n) = \frac{n}{2} + \frac{n - 2}{2} + \frac{n-2}{2} + \frac{n - 4}{2} + \frac{n - 4}{2} + \frac{n - 6}{2} + \frac{n - 6}{2} + ...+ 2 + 2 + 1 + 1.  
\end{equation}
Thus, one has
\begin{eqnarray}
D_{A,max}(n) & = & \frac{n}{2} + 2 \sum_{i = 1}^{\frac{n - 2}{2}} i \\
             & = & \frac{n^2}{4}.
\end{eqnarray}
This allows one to conclude that 
\begin{equation}
D_{A,max} =  \frac{n^2 - x}{4},
\end{equation}
where $x$ is a binary variable indicating if $n$ is odd ($x = 1$ if $n$ is odd; $x = 0$ otherwise).
Interestingly, $D_{A,max}$ coincides with $D_{min}^{star}$, the value of $D_{min}$ of a star tree defined in (\ref{mla_star_tree_equation}). Since $D_{min} \leq D_A$ and $D_A \leq D_{min}^{star}$ we conclude that 
$D_{min} \leq D_{min}^{star}$ with equality if the tree is a star tree. See Appendix A for details on how we validated this result.

It is easy to prove that $D_{min}^{star} \leq D_{random}$. 
By the definitions of $D_{min}^{star}$ and $D_{random}$ (equations (\ref{mla_star_tree_equation}) and (\ref{rla_scaling_of_lengths_equation}), respectively), this is equivalent to  
\begin{equation}
\frac{n^2-x}{4} \leq \frac{n^2 - 1}{3},
\end{equation}
which becomes
\begin{equation}
4 - 3x \leq n^2
\end{equation}
after some algebra. Recalling that $x$ is indeed a function of $n$, a simple evaluation of the inequality from $n = 1$ onwards allows one to conclude that $D_{min}^{star} \leq D_{random}$ holds for $n \geq 1$, with equality if and only if $n = 1$ or $n = 2$.

\section{The scaling of $D_{min}$ in uniformly random labelled trees}

\label{scaling_section}

To investigate the scaling of $\left< D_{min} \right>$ in uniformly random labelled trees, we generated random labelled trees and calculated the value of $D_{min}$ for each tree using Shiloach's algorithm \cite{Shiloach1979}. Since algorithms of this kind have remained theoretical for decades (they have not been implemented and used in depth) a thorough testing of our implementation of Shiloach's algorithm is vital. See the Appendix for details about the tests that we considered to validate that implementation. 

A uniformly random labelled tree can be generated in different ways. One possibility is the Aldous-Brother algorithm \cite{Aldous1990a,Broder1989a}, assuming a complete graph as the basis of the random walk. Another possibility is to generate a uniformly random Pr\"ufer code and then to obtain the corresponding tree. A Pr\"ufer code for a tree of $n$ nodes is a sequence of $n - 2$ integers between $1$ and $n$ that identifies a unique labelled tree \cite{Pruefer1918a}. We decided to use Pr\"ufer codes for generating random trees because the same procedure is also helpful to generate all possible labelled trees when testing Shiloach's algorithm (see Appendix).   

\begin{figure}
\begin{indented}
\item[]
\includegraphics[scale = 0.7]{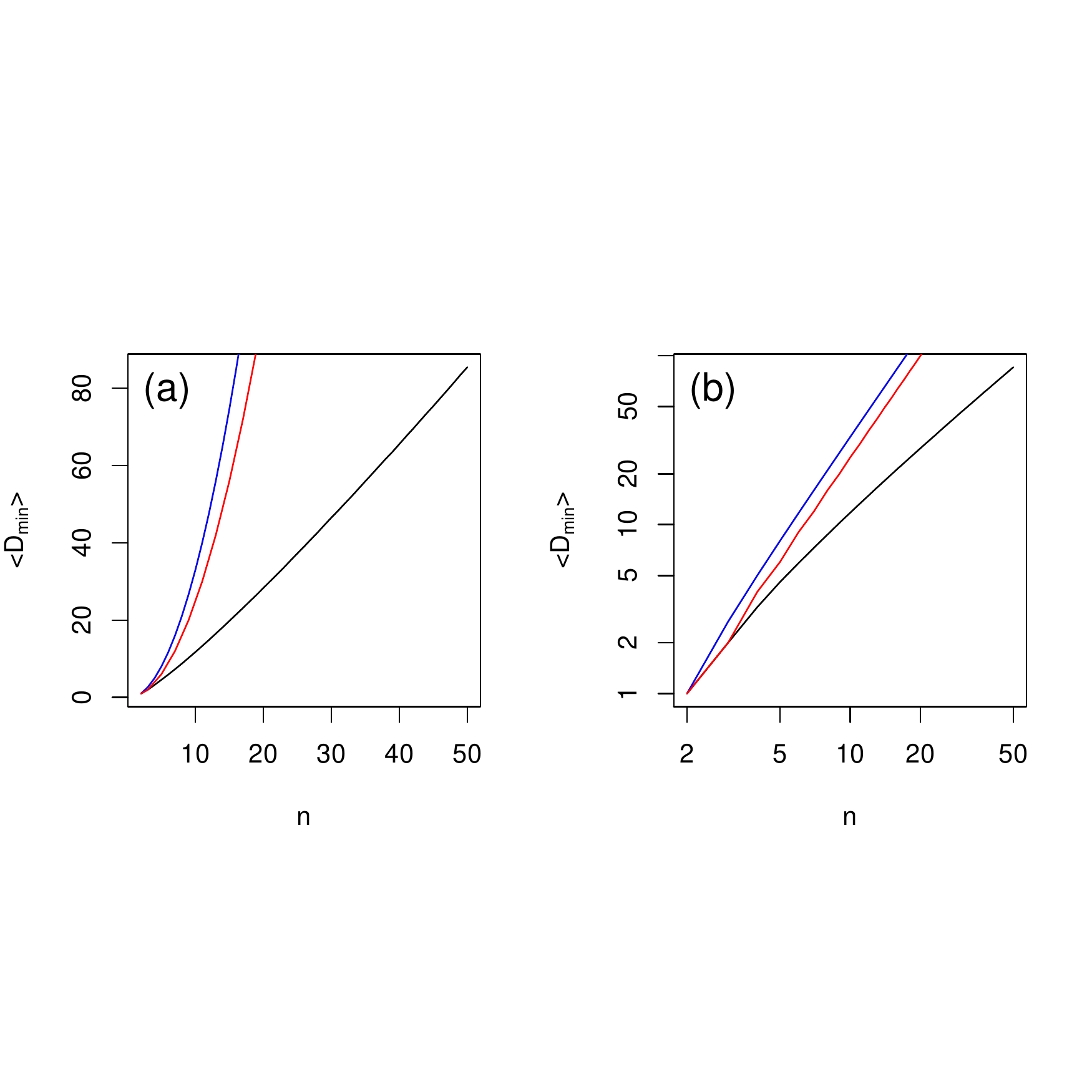}
\caption{\label{sum_of_dependency_lengths_figure} (a) The growth of $\left<D_{min}\right>$, the average minimum sum of dependency lengths in uniformly random trees, as a function of $n$, the number of vertices of the tree (black line). For a given tree size, $\left<D_{min}\right>$ is estimated over $10^4$ uniformly random trees. For reference, we also show $D_{min}^{star}$, the upper bound of $D_{min}$ (red line), $D_{random}$, the expected value of the sum of dependency lengths in uniformly random linear arrangements (blue line) and $\left< D \right>$, 
the mean value of $D$ in uniformly random labelled trees where vertex labels are taken as vertex positions (yellow line). The latter cannot be seen because it is covered by the curve of $D_{random}$.
 (b) The same as (a) in double logarithmic scale.   }
\end{indented}
\end{figure}

Figure \ref{sum_of_dependency_lengths_figure} shows that the growth of $\left<D_{min}\right>$ as a function of $n$ is almost linear in uniformly random labelled trees. Figure \ref{sum_of_dependency_lengths_figure} also shows the equivalence between $D_{random}$, the expected value of $D$ in a uniformly random linear arrangement, and $\left< D \right>$ as defined above.

\begin{figure}
\begin{indented}
\item[]
\includegraphics[scale = 0.7]{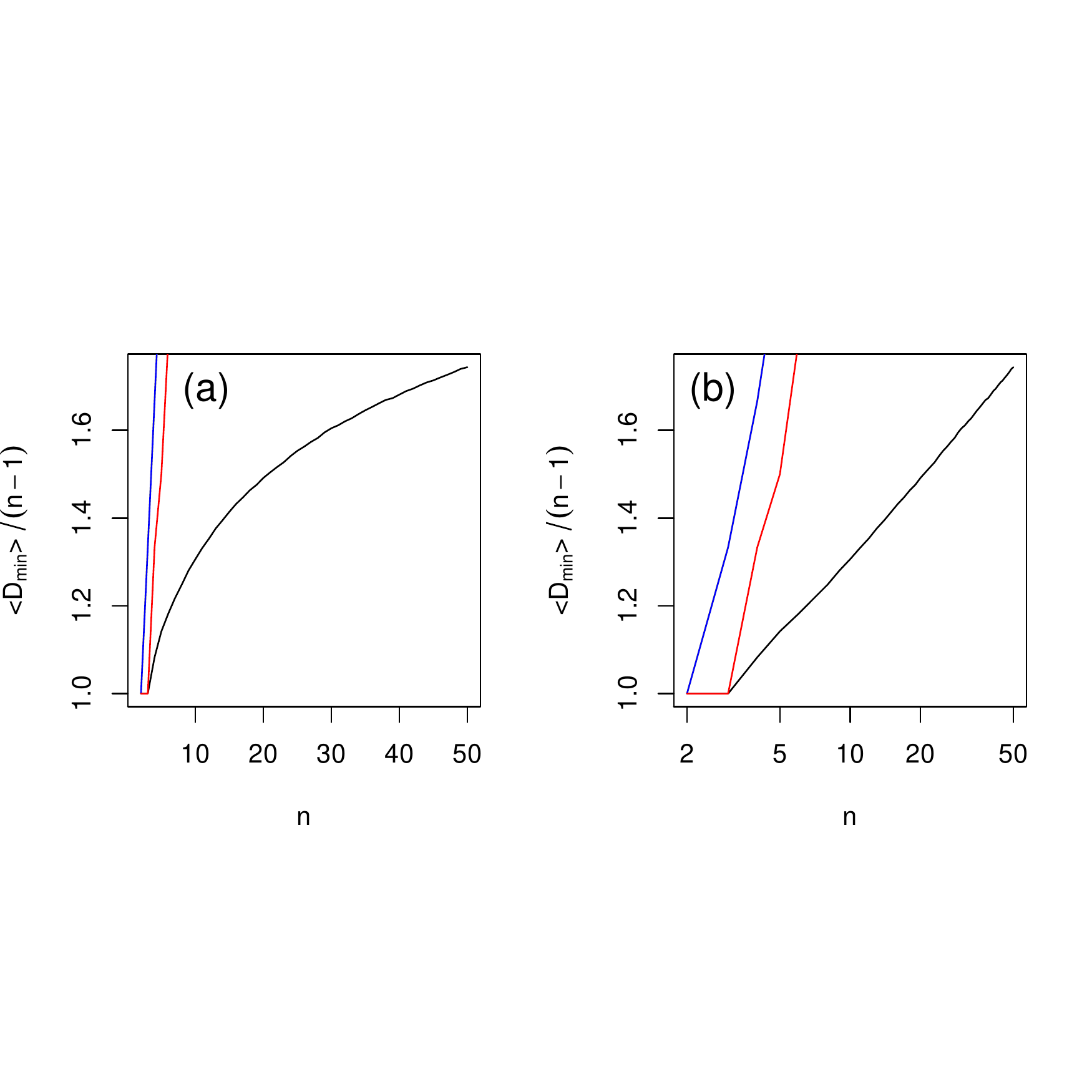}
\caption{\label{mean_dependency_length_figure} (a) The growth of $\left< D_{min} \right>/(n-1)$, the average minimum mean dependency length in uniformly random trees, as a function of $n$, the number of vertices of the tree (black line). $\left< D_{min} \right>$ is estimated over $10^4$ uniformly random trees.
For reference, we also show $D_{min}^{star}/(n-1)$, the upper bound of $D_{min}/(n-1)$ (red line) and $D_{random}/(n-1)$, the expected value of the mean dependency length in uniformly random linear arrangements (blue line). (b) The same as (a) but using logarithmic scale for the $x$-axis.}
\end{indented}
\end{figure}

$\left<D_{min}\right>/(n-1)$, i.e. the mean dependency length of minimum linear arrangements of uniformly random trees of $n$ vertices, will help us to shed light on the actual dependency between $D_{min}$ and $n$. Figure \ref{mean_dependency_length_figure} (a) suggests that $\left<D_{min}\right>/(n-1)$ grows logarithmically as $n$ increases. Such a logarithmic growth is confirmed by the straight line that appears when taking logs on the $x$-axis for $n \geq 3$ (figure \ref{mean_dependency_length_figure} (b)). Notice that the value of $\left<D_{min}\right>/(n-1)$ is the same for $n = 2$ and $n = 3$. 

\begin{figure}
\begin{indented}
\item[]
\includegraphics[scale = 0.5]{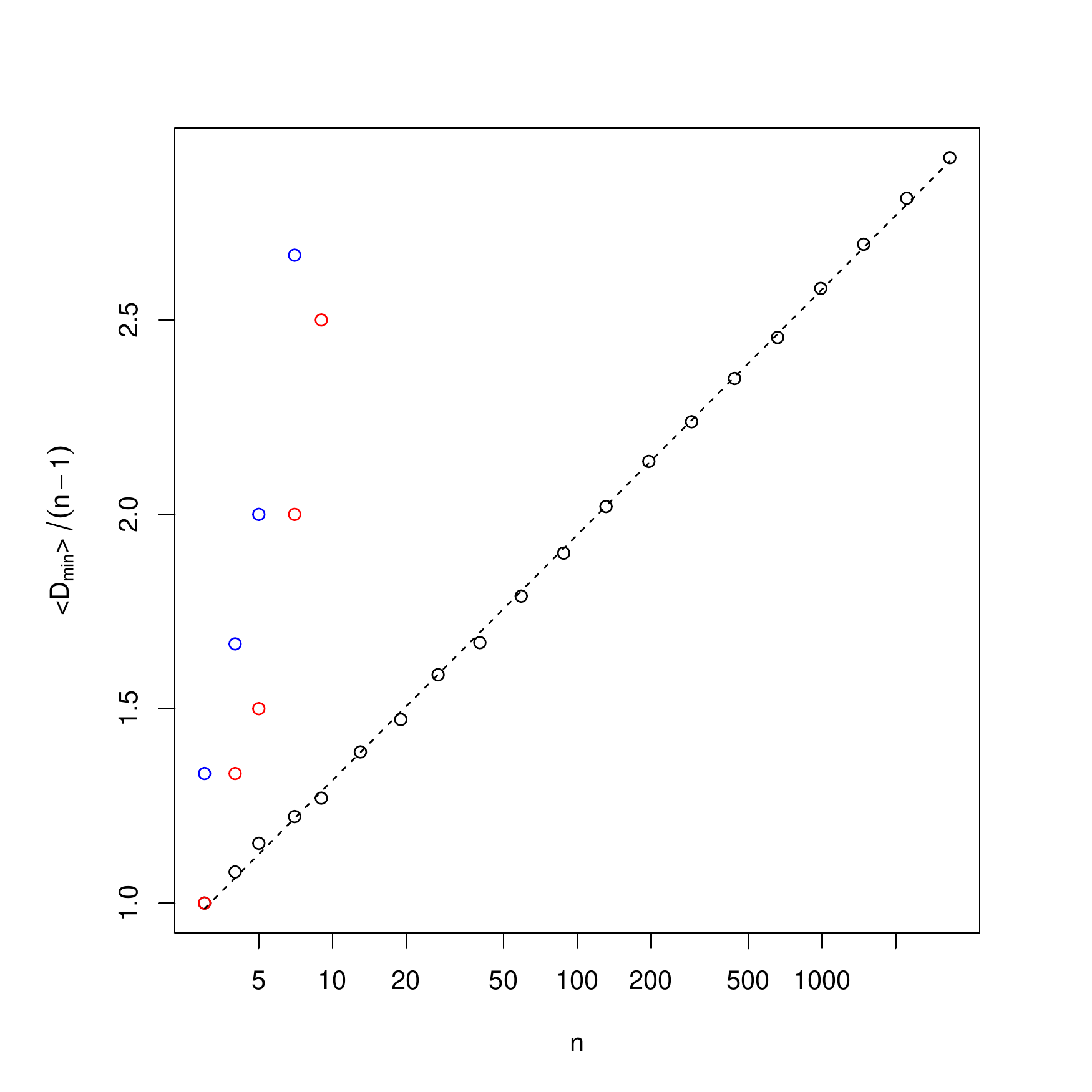}
\caption{\label{scaling_of_mean_dependency_length_figure} The growth of $\left<D_{min}\right>/(n-1)$, the average minimum mean dependency length in uniformly random trees, as a function of $n$, the number of vertices of the tree (black circles). $\left<D_{min}\right>$ is estimated over 200 uniformly random trees. 
The values of $n$ chosen are obtained with $n = \lfloor (3/2)^{k} + 2 \rfloor$ for $k = 1,2,3,...,20$. For reference, we also show the best fit of $D_{min}/(n-1) = a \log n + b$ (dashed line),
$D_{min}^{star}/(n-1)$, the upper bound of $D_{min}/(n-1)$ (red circles), and $D_{random}/(n-1)$, the expected value of the mean dependency lengths in uniformly random linear arrangements (blue circles).}
\end{indented}
\end{figure}

The logarithmic growth of $\left<D_{min}\right>/(n-1)$ is in stark contrast to 
\begin{itemize}
\item
The linear growth of $\left<D\right>/(n-1)$ in random linear arrangements (or equivalently, as shown above, in uniformly random labelled trees where vertex labels are taken as vertex positions), as expected from (\ref{rla_scaling_of_lengths_equation}), which gives $\left<D\right>/(n-1) = (n+1)/3$ (figure \ref{mean_dependency_length_figure}). 
\item
The linear growth of the upper bound of $\left<D_{min}\right>/(n-1)$, i.e. $D_{min}^{star}/(n-1)$ which is $\approx (n+1)/4$ according to (\ref{mla_star_tree_equation}) (Fig. \ref{mean_dependency_length_figure}). 
\end{itemize}
Further support for the logarithmic growth of $\left<D_{min}\right>/(n-1)$ is provided in figure \ref{scaling_of_mean_dependency_length_figure},
where exponentially increasing values of $n$ are employed to check if the growth is the same for large values of $n$. Again a straight line is recovered when logs are taken on the $x$-axis. Interestingly, a least squares linear regression for the relationship between $\left<D_{min}\right>/(n-1)$ and $\log n$ for $n \geq 3$ in figure \ref{scaling_of_mean_dependency_length_figure} provides support for   
\begin{equation}
D_{min}/(n-1) \approx a \log n + b
\label{mean_dependency_length_versus_vertices_equation}
\end{equation}
with $a = 0.27$ and $b = 0.68$. Figure \ref{scaling_of_mean_dependency_length_figure} indicates that (\ref{mean_dependency_length_versus_vertices_equation}) predicts the true values with high accuracy. $n = 2$ is excluded from the fit because $D_{min}/(n-1) = 1$ for both $n = 2$ and $n = 3$ and the function that we are fitting is strictly monotonous.  
Equation (\ref{mean_dependency_length_versus_vertices_equation}) allows one to conclude that $D_{min}$ follows 
\begin{equation}
\left<D_{min} \right> \approx a (n-1) \log n + (n - 1) b.
\end{equation}
with high accuracy. 

\section{Discussion}

\label{discussion_section}

In this article, we have improved our understanding of the limits of the variation of $D_{min}$ in trees. 
The results presented in Section \ref{upper_bound_section} allow one to conclude that 
\begin{equation}
D_{min}^{linear} \leq D_{min} \leq D_{min}^{star} \leq D_{random} 
\label{variation_mla_equation}
\end{equation}
for $n \geq 1$.
The bounds of $D_{min}$ involving linear and star trees are reminiscent of the limits of the variation of $\left<k^2 \right>$, the second moment of degree about zero of a network of $n$ vertices, i.e. 
\begin{equation}
\left< k^2 \right> = \frac{1}{n} \sum_{i=1}^n k_i^2,
\end{equation}
where $k_i$ is the degree of the $i$-th vertex. Interestingly, $\left<k^2 \right>$ in trees of the same size obeys \cite{Ferrer2013b}
\begin{equation}
\left<k^2 \right>^{linear} \leq \left<k^2 \right> \leq \left<k^2 \right>^{star}, 
\end{equation}
where $\left<k^2 \right>^{linear}= 4 - 6/n$ is the value of $\left<k^2 \right>$ in a linear tree and 
$\left<k^2 \right>^{star}= n - 1$ is the value of $\left<k^2 \right>$ in a star tree.

In this article, we have also shown that the mean edge length in optimal linear arrangements grows logarithmically with the size of the tree. The origins of such a growth should be the subject of future research.
Note that (\ref{variation_mla_equation}) and the definitions of $D_{min}^{linear}$ and $D_{min}^{star}$ in (\ref{mla_linear_tree_equation}) and (\ref{mla_star_tree_equation}) imply that
\begin{equation}
1 \leq \frac{D_{min}}{n-1} \leq \frac{n^2}{4(n-1)} \approx \frac{n}{4}.
\end{equation}
Given this wide range of variation, the fact that $\left<D_{min}\right>/(n - 1)$ grows logarithmically with $n$ suggests that $\left<D_{min}\right>/(n - 1)$
is dominated by trees with low $D_{min}$ far from the linear growth of star trees and closer to linear trees. A similar behavior is found in the scaling of the expected number of edge crossings in uniformly random labelled trees, which is far from that of star trees and closer to that of linear trees \cite{Ferrer2013d}. 
The origins of that logarithmic growth should be the subject of future research. A possible application of our result could be in research on the scaling of $D$ or $D/(n - 1)$ in syntactic dependency trees, where pressure to minimize dependency lengths is supported both empirically \cite{Ferrer2004b,Liu2008a,Ferrer2013c,Futrell2015a} and theoretically \cite{Ferrer2008e,Ferrer2013e,Ferrer2014f}. The logarithmic dependency described by equation (\ref{scaling_of_mean_mla_equation}) appears to be a relevant candidate model for the actual dependency between $D/(n - 1)$ and $n$ in those trees \cite{Ferrer2004b,Futrell2015a}. The suitability of this candidate may depend on the extent to which real syntactic dependency lengths are optimized. Having said this, the logarithmic dependency is an unavoidable candidate to investigate the scaling of the actual dependency between $\left<D_{min}\right>/(n - 1)$ and $n$ in optimal linear arrangements of syntactic dependency trees and other kinds of real trees.

\appendix

\section{Validation of Shiloach's algorithm}

To solve the minimum linear arrangement problem we implemented Shiloach's algorithm incorporating a recent correction \cite{Esteban2015a}.

For a given tree, the correctness of the value of $D_{min}$ calculated by our implementation of the corrected version of Shiloach's algorithm for a given tree was checked in three different ways: 
\begin{enumerate}
\item
Since the output of Shiloach's algorithm is both $D_{min}$ and $\pi_{min}$, a minimum linear arrangement (a one-to-one mapping $\pi$ yielding $D_{min}$), 
we checked that the value of $D_{min}$ coincides with the value of $D$ obtained from $\pi_{min}$. 
\item
By means of known examples or theoretical results giving the exact value of $D_{min}$ or bounds.  
\item
By means of a brute force algorithm that allows one to check the correctness of the results for small trees. The algorithm is less error prone than Shiloach's algorithm, as it is conceptually simpler and much easier to implement,  
but it is computationally very expensive.
\end{enumerate}
The next subsections provide further details about the second and the third evaluation procedure. 

For a given $n$, we performed two kinds of exploration of the space of possible trees: 
\begin{itemize}
\item
An exhaustive exploration, i.e. Shiloach's algorithm was tested against all possible labelled trees, generated with the help of Pr\"ufer codes \cite{Pruefer1918a}.
\item
Random exploration, i.e. Shiloach's algorithm was tested against uniformly random trees, obtained by generating uniformly random Pr\"ufer codes \cite{Pruefer1918a}.
\end{itemize} 
Given the high cost of the brute force algorithm, we performed explorations with and without using the brute force test, in order to be able to test Shiloach's algorithm with larger trees.
All these options lead to four possibilities for validation that are summarized in Table \ref{validation_table} with the corresponding values of $n$ that were used in each case.
These tests are not only used to check the correctness of our implementation of Shiloach's algorithm, but also serve as a test for the inequality $D_{min} \leq D_{min}^{star}$, derived in Section \ref{upper_bound_section}. 

\begin{table}
\caption{\label{validation_table} A summary of the values of $n$ that were used for testing in all the four evaluation conditions. The number of random samples used varies for computational reasons: 20 for *, 100 for ** and 200 for ***, as it is explained in figure \ref{scaling_of_mean_dependency_length_figure}.}
\begin{indented}
\item[]\begin{tabular}{ccc}
\br
                & all tests           & all tests excluding the brute force algorithm \\
\mr
exhaustive      & $1 \leq n \leq 10$  & $1 \leq n \leq 11$ \\
random sampling & $11 \leq n \leq 19^*$ & $12 \leq n \leq 1000^{**}$ \\
                &                     & $n = (3/2)^k + 2$ for $k = 1,2,...,20^{***}$ \\
\br
\end{tabular}
\end{indented}
\end{table} 
 
\subsection{Exact values or bounds for $D_{min}$}

Our implementation of Shiloach's algorithm was tested with trees for which $D_{min}$ can be obtained via formulae:
\begin{itemize}
\item
Linear trees (recall (\ref{mla_linear_tree_equation})).
\item 
Star trees (recall (\ref{mla_star_tree_equation})).
\item
Complete binary trees (recall (\ref{mla_complete_binary_tree_equation})).
When $k = 5$, $D_{min} = 60$. This means that on average, the length of an edge is two (since $k = 5$ implies $n = 30$ vertices).
Complete binary trees are powerful test cases for two reasons:
  \begin{itemize}
  \item
  For $k<5$ the m.l.a. is given by an inorder traversal of the vertices starting on the root of the binary tree. For $k\geq 5$ the strategy ceases to provide the solution of the m.l.a. \cite{Chung1978a}.
  \item
  The original version of Shiloach's algorithm fails when $k \geq 5$ \cite{Esteban2015a}.
  \end{itemize}
\item
A kind of ternary trees (recall (\ref{mla_kind_of_ternary_tree_equation})). 
\end{itemize}

\begin{figure}
\begin{indented}
\item[]
\includegraphics[scale = 0.75]{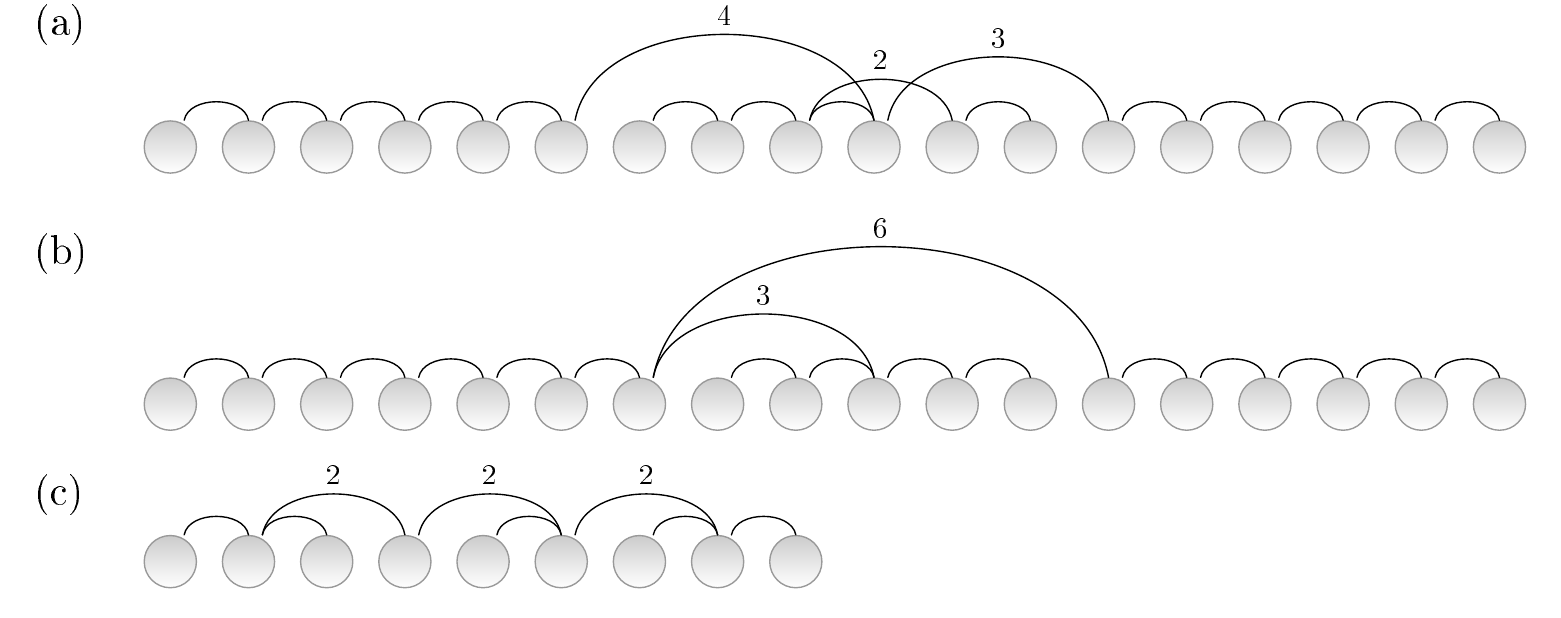}
\caption{\label{examples_of_mla_figure} Minimum linear arrangements of trees (only the length of edges that are longer than unity is indicated) (a) The minimum linear arrangement of a tree. The total sum of dependency lengths is $D = 14 \cdot 1 + 1 \cdot 2 + 1 \cdot 3 + 
1 \cdot 4 = 23$. (b) A minimum linear arrangement of the same tree of (a) when crossings are disallowed, with $D = 15 \cdot 1 + 1 \cdot 3 + 1 \cdot 6 = 24$.  
(c) A minimum linear arrangement of a syntactic dependency tree with $D = 5 \cdot 1 + 3 \cdot 2 = 11$. 
(a) and (b) are adapted from \cite{Hochberg2003a}. (c) is adapted from \cite{Baronchelli2013a}. }
\end{indented}
\end{figure}

\begin{figure}
\begin{indented}
\item[]
\includegraphics[scale = 0.75]{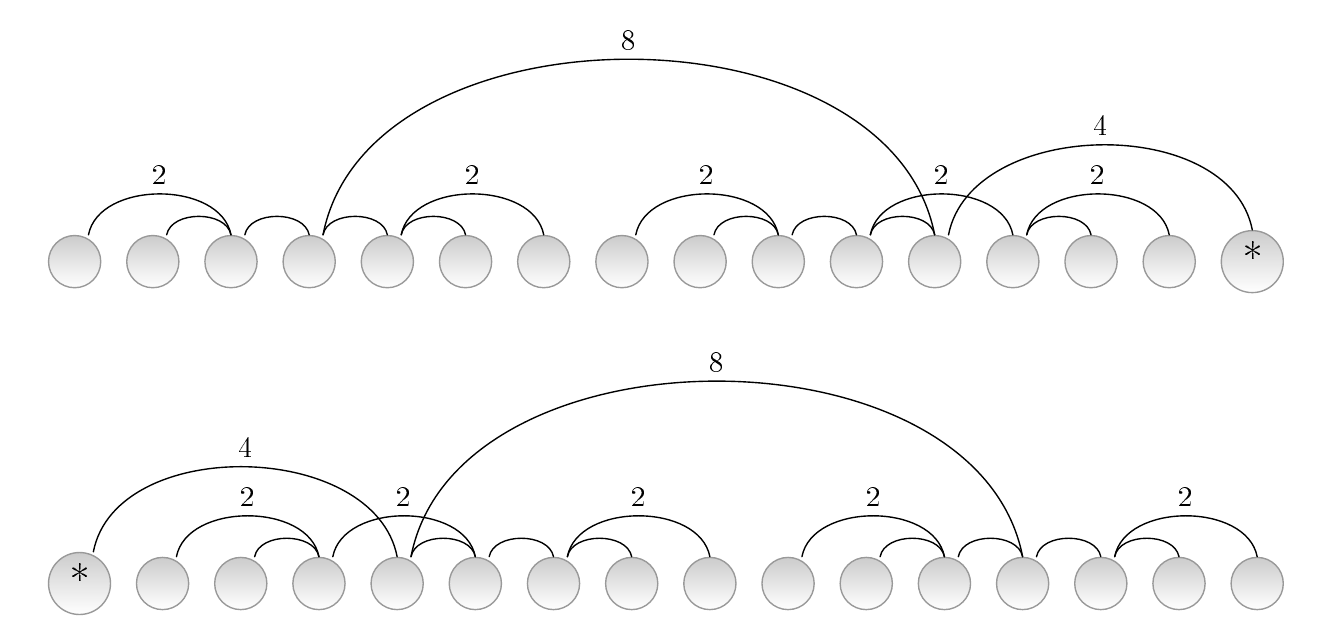}
\caption{\label{complete_tree}  A minimum linear arrangement of a complete binary tree of $5$ levels with $D = 16 \cdot 1 + 10 \cdot 2 + 2 \cdot 4 + 2 \cdot 8 = 60$. The example is adapted from \cite{Chung1978a}. The tree is too long, therefore it is broken into two lines. The vertex marked with * is shared by the two pieces of the tree. }
\end{indented}
\end{figure}

The solution of the m.l.a. for concrete trees is shown in previous publications:
\begin{itemize}
\item
Suppose that $D_{min}^{NC}$ is the solution to the m.l.a. problem when edge crossings are not allowed. Figure 1 of \cite{Hochberg2003a} is an interesting test because the solution to the m.l.a. when crossings are not allowed ($D_{min}^{NC} = 24$) differs from the solution of the unconstrained m.l.a. ($D_{min} = 23$). These examples are reproduced in figures \ref{examples_of_mla_figure} (a-b).
\item
Another example of a minimum linear arrangement with crossings is the complete $5$-level binary tree in figure 1 of \cite{Chung1978a}. 
The linear arrangement in that figure has a typo: the vertex labelled with 12 has two successors: a vertex labelled with 4 and another labelled with 1. The label of the latter should be 11. Figure \ref{complete_tree} shows the correct linear arrangement. 
\item
Figure 4C of \cite{Baronchelli2013a} with $D_{min} = 11$ (figure \ref{examples_of_mla_figure} (c)).
\end{itemize}

In general, the value of $D_{min}$ must satisfy the following properties:
\begin{itemize}
\item
$D_{min} \leq D$, where $D$ is the actual sum of dependency lengths of the tree or the sum of dependency lengths that is obtained interpreting vertex labels between $1$ and $n$ as vertex positions.  
\item
$D_{min} \leq D_{min}^{star}$, with $D_{min}^{star}$ defined as in (\ref{mla_star_tree_equation}).
\item
$D_{min} \leq D_{min}^{NC}$. $D_{min}^{NC}$ is calculated in linear time with Hochberg \& Stallmann's linear time algorithm \cite{Hochberg2002a,Hochberg2003a}.
\item
$D_{min}$ is bounded below by a function of $n$ and $\left< k^2 \right>$ (the degree second moment about zero of a tree) as \cite{Ferrer2013b}
\begin{equation}
D_{min} \geq \frac{n}{8(n-1)}\left< k^2 \right> + \frac{1}{2}.
\end{equation} 
\end{itemize}

\subsection{Validation with the help of a brute force algorithm}

The results of our implementation of Shiloach's algorithm to solve the m.l.a. problem are compared against those of a brute force algorithm for small trees. 
Tentatively, the brute force algorithm should be simpler and therefore less error prone. 
A straightforward brute force algorithm consists of generating the $n!$ permutations of the vertices to find the smallest $D$. This huge permutation space is reduced a little bit noting that all the permutations where the leaves attached to the same internal vertex have exchanged their positions give the same $D$. If the space of permutations is explored in a way that all the leaves attached to the same internal vertex are visited always in the same order, $S$, the size of the space to explore, reduces from $S = n!$ to 
\begin{equation}
S = \frac{n!}{\Pi_{i=1}^n (l_i!)},
\end{equation}  
where $l_i = 1$ if the $i$-th vertex is not connected to any leaf (this can happen if the $i$-th vertex is a leaf or the $i$-th vertex is an internal vertex that it is not connected to any leaf); otherwise, $l_i$ is the number of leaves attached to the $i$-th vertex. Suppose a star tree. Then $l_i = n - 1$ for the hub and $l_i = 1$ for the remainder of the vertices and then $S = n$ (a dramatic reduction of the space of permutations). Suppose a linear tree, then $l_i = 1$ for every vertex and then $S = n!$ (no reduction of the space of permutations). 
 



\ack
We are grateful to A. D\'iaz-Guilera, O. Sagarra and C. P\'erez-Vicente for helpful comments and discussions.
JLE is funded by the project TASSAT2 (TIN2013-48031-C4-1-P) from MINECO (Ministerio de Economia y Competitividad). 
RFC is funded by the grants 2014SGR 890 (MACDA) from AGAUR (Generalitat de Catalunya) and also
the APCOM project (TIN2014-57226-P) from MINECO.
CGR is funded by the TELEPARES-UDC project (FFI2014-51978-C2-2-R) from MINECO, the grant R2014/034 from Xunta de Galicia, and the Oportunius program from the Galician Innovation Agency (Xunta de Galicia).

\section*{References}

\bibliographystyle{unsrt}

\bibliography{../../TACL-Crossings/main,../../TACL-Crossings/twoplanaracl,../../TACL-Crossings/Ramon}

\end{document}